\documentclass[%
reprint,
superscriptaddress,
footinbib,
twocolumn,
amsmath,amssymb,
aps,
prl,
floatfix,
longbibliography
]{revtex4-1}

\usepackage{amsmath}
\usepackage{graphicx}
\usepackage{dcolumn}
\usepackage{bm}
\usepackage{amssymb}
\usepackage{multirow}
\usepackage[colorlinks=true,citecolor=blue]{hyperref}

\begin{document}

\title{Realistic Spin Model for Multiferroic NiI$_2$}

\author{Xuanyi Li}
\affiliation{Key Laboratory of Computational Physical Sciences (Ministry of Education), Institute of Computational Physical Sciences, State Key Laboratory of Surface Physics, and Department of Physics, Fudan University, Shanghai 200433, China.}

\author{Changsong Xu}
\email{csxu@fudan.edu.cn}
\affiliation{Key Laboratory of Computational Physical Sciences (Ministry of Education), Institute of Computational Physical Sciences, State Key Laboratory of Surface Physics, and Department of Physics, Fudan University, Shanghai 200433, China.}
\affiliation{Shanghai Qi Zhi Institute, Shanghai 200030, China}

\author{Boyu Liu}
\affiliation{Key Laboratory of Computational Physical Sciences (Ministry of Education), Institute of Computational Physical Sciences, State Key Laboratory of Surface Physics, and Department of Physics, Fudan University, Shanghai 200433, China.}

\author{Xueyang Li}
\affiliation{Key Laboratory of Computational Physical Sciences (Ministry of Education), Institute of Computational Physical Sciences, State Key Laboratory of Surface Physics, and Department of Physics, Fudan University, Shanghai 200433, China.}

\author{L. Bellaiche}
\affiliation{Physics Department and Institute for Nanoscience and Engineering, University of Arkansas, Fayetteville, Arkansas 72701, USA}%

\author{Hongjun Xiang}
\email{hxiang@fudan.edu.cn}
\affiliation{Key Laboratory of Computational Physical Sciences (Ministry of Education), Institute of Computational Physical Sciences, State Key Laboratory of Surface Physics, and Department of Physics, Fudan University, Shanghai 200433, China.}
\affiliation{Shanghai Qi Zhi Institute, Shanghai 200030, China}


\begin{abstract}
  A realistic first-principle-based spin Hamiltonian is constructed for the type-II multiferroic NiI$_2$, using a symmetry-adapted cluster expansion method.
  Besides single ion anisotropy and isotropic Heisenberg terms, this model further includes the Kitaev interaction and a biquadratic term, and
  can well reproduce striking features of the experimental helical ground state, that are, {\it e.g.}, a proper screw state, canting of rotation plane, propagation direction and period.  Using this model to build a phase diagram, it is demonstrated that, (i) the in-plane propagation direction of $\langle1\bar10\rangle$ is determined by the Kitaev interaction, instead of the long-believed exchange frustrations; and (ii) the canting of rotation plane is also dominantly determined by Kitaev interaction, rather than interlayer couplings.
  Furthermore, additional Monte Carlo simulations  reveal three equivalent domains and different topological defects.
  Since the ferroelectricity is induced by spins in type-II multiferroics, our work also implies that Kitaev interaction is closely related to the multiferroicity of NiI$_2$.
\end{abstract}
\maketitle

The materials of van der Waals type can potentially be made into two-dimensional (2D) layers, which exhibit exceptional properties, such as massless fermions \cite{novoselov2005two}, valleytronics \cite{mak2012control}, ferroelectricity \cite{chang2016discovery} and ferromagnetism \cite{gong2017discovery,huang2017layer}.
Recently, electromagnetic couplings were observed in few layers and monolayers of NiI$_2$, which makes NiI$_2$ the first established 2D multiferroic \cite{ju2021possible,song2022evidence}.


Bulk NiI$_2$ has been known as a van der Walls layered type-II multiferroic. It crystallizes in a rhombohedral lattice with a space group of $R\bar{3}m$ (point group $D_{3d}$). Each layer of NiI$_2$ consists of edge-sharing NiI$_6$ octahedra, yielding a triangular lattice of magnetic Ni$^{2+}$ ions, as shown in Fig. 1a. The Ni$^{2+}$ ion exhibits an electronic configuration of $3d^8$, with fully occupied $t_{2g}$ orbits and half-filled $e_g$ orbits, resulting in the spin value of $S=1$ and a local moment of 2 $\mu_B$ on each Ni$^{2+}$.
The ground state was determined to be a proper screw (PS) state, where spins rotate in a plane that is perpendicular to the propagation direction.
This proper screw is characterized by ${\bm q}\approx(0.138, 0, 1.457)$ in the bulk system \cite{kuindersma1981magnetic}, which indicates in-plane propagation along ${\langle1\bar10\rangle}$ directions with a period of $\lambda\approx7.23{\bm a}$, and the out-of-plane propagation arises from the interlayer antiferromagnetic (AFM) alignments.
As NiI$_2$ is insulating, such PS state breaks the inversion symmetry and induces an electric polarization along  ${\langle110\rangle}$ directions \cite{kuindersma1981magnetic,kurumaji2013magnetoelectric}.

To understand the specific propagation directions of the PS of NiI$_2$,
analytical results on $J_1-J_2-J_3$ model of triangular lattice indicate that (i) the ${\langle1\bar10\rangle}$ propagation can be stabilized by FM $J_1$ and AFM $J_2$ with $J_2/J_1<-1/3$; while (ii) the ${\langle110\rangle}$ propagation is favored by FM $J_1$ and AFM $J_3$ with $J_3/J_1<-1/4$ \cite{okubo2012multiple}.
However, various models extracted from density functional theory (DFT) actually predict a ${\langle110\rangle}$ propagated ground state, with $J_1$ and $J_2$ both being FM \cite{amoroso2020spontaneous,riedl2022microscopic,ni2021giant,ni2022plane}, implying that the competing $J_1-J_2$ mechanism is not suitable for NiI$_2$.
Moreover, even though the Heisenberg model can stabilize a ${\langle1\bar10\rangle}$ propagation, it can not explain why the ground state is PS, instead of other degenerate helical states (see Figs. 1c and 1d).


Another interesting but still elusive point is the canting of the spin rotation plane. Measurements find that the normal of the rotation plane is not along the in-plane  ${\langle1\bar10\rangle}$ propagation direction, but rather forms an angle of 55$^\circ$ with the out-of-plane direction of NiI$_2$ bulk  \cite{kuindersma1981magnetic}. Such canting has been believed to be natural, as the presumed PS state should have its rotation plane being perpendicular to its propagation direction and the PS state of NiI$_2$ does have an out-of-plane propagation component \cite{kurumaji2011magnetic,kurumaji2013magnetoelectric}.
However, common mechanisms can not explain such canting, as (i) single ion anisotropy (SIA) does not favor specific canting angle; (ii) the Dzyaloshinskii-Moriya interaction (DMI)
is not allowed by the inversion symmetry of NiI$_2$ (Note that incommensurate spin patterns are too weak to generate non-negligible DMI);
and (iii) interlayer Heisenberg terms are proved to have effects neither on propagation directions nor cantings \cite{regnault1982inelastic}.
On the other hand, new forms of interactions, {\it i.e.}, Kitaev interaction \cite{stavropoulos2019microscopic,amoroso2020spontaneous,riedl2022microscopic} and  biquadratic interactions \cite{ni2021giant}, have recently been proposed to be non-negligible in NiI$_2$, but their effects and interplays are still not clearly understood.
Hence, any highly desired realistic model of NiI$_2$ has to not only incorporate all these aforementioned important mechanisms, but also reproduce the correct ground state -- which is currently lacking.

In this work, we build a first-principle-based spin Hamiltonian for NiI$_2$, taking advantage  of a symmetry-adapted cluster expansion and machine learning methods.
The resulting Hamiltonian can well reproduce the observed PS state of NiI$_2$, with the propagation, period and canting angle comparing well with experiments on bulk systems.
By further developing a phase diagram, it is demonstrated that (i) Heisenberg terms actually lead to ${\langle110\rangle}$ propagation; and (ii) it is the Kitaev interaction that not only results in the actual ${\langle1\bar10\rangle}$ propagation, but also dominantly determines the canting of the rotation plane.
The roles of biquadratic interaction and interlayer couplings are also carefully examined. Monte Carlo (MC) simulations further predict diverse spin textures and topological defects.

\begin{figure}[t]
  \centering
  \includegraphics[width=8cm]{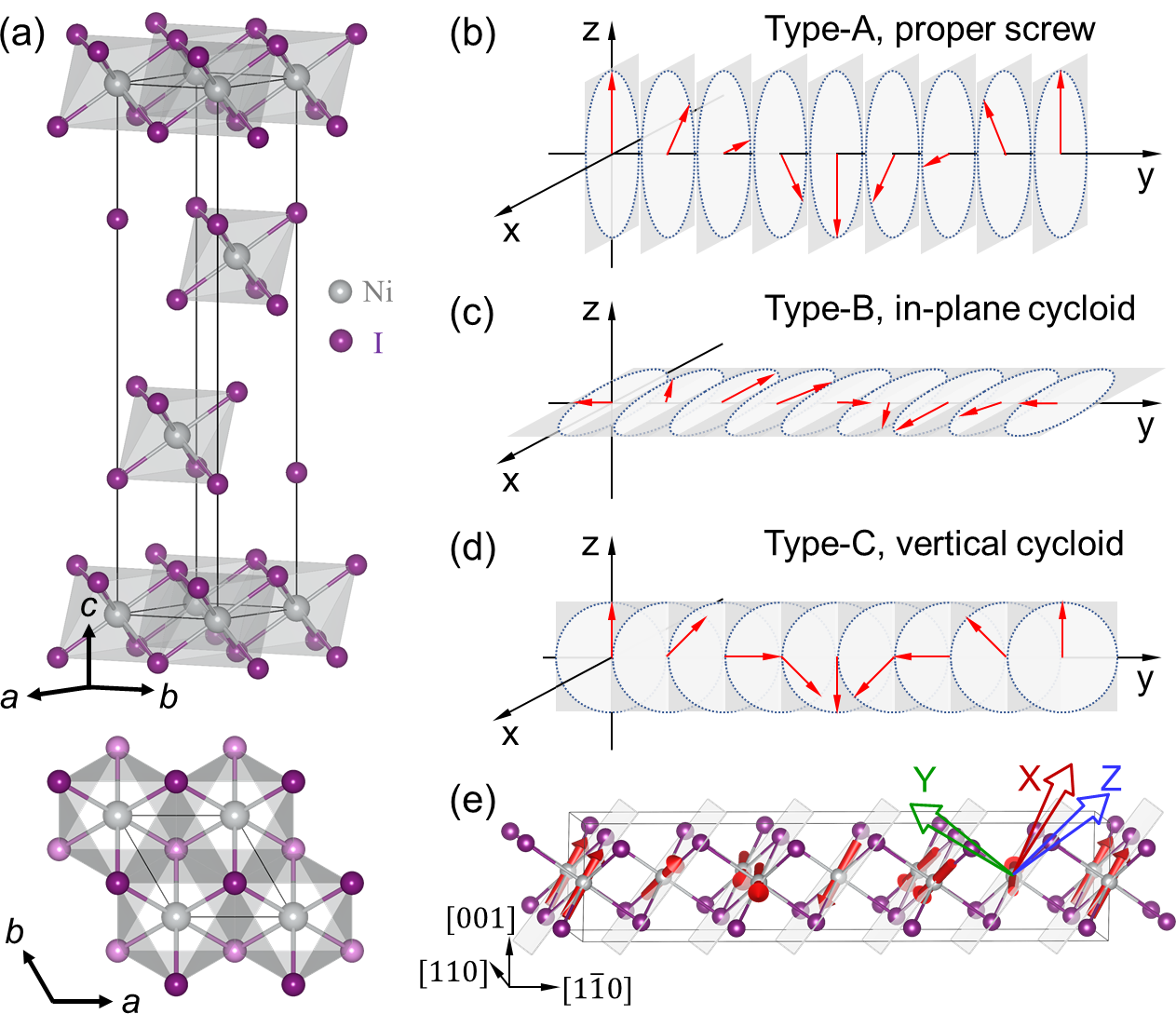}
  \caption{Schematics of (a) NiI$_2$ crystal structure and common helical spin structures, (b) proper screw, (c) in-plane cycloid and (d) vertical cycloid. Panel (e) displays the PS$^{\langle1\bar10\rangle}_{cant}$ state of NiI$_2$, where spins rotate in a canted plane that is spanned by the Ni$_2$I$_2$ clusters. The hollow red, green and blue arrows denote the Kitaev basis $\{XYZ\}$.}
\end{figure}

Our newly developed symmetry-adapted cluster expansion method, as implemented in the PASP software, is applied to build the spin Hamiltonian  of  NiI$_2$ \cite{lou2021pasp,xu2022assembling}. Such method roots in cluster expansion that goes over all combinations of spin components of $S_{\alpha} (\alpha=x,y,z)$. By further applying crystal symmetries to those combinations, only the symmetry-allowed terms, {\it i.e.}, the invariants, are kept.
The coefficients of these invariants can be fitted from total energies obtained from DFT calculations via a machine learning algorithm \cite{li2020constructing}. Such method can thus in principle consider all possible interactions to any body and any order [see Supplemental Materials (SM) for details \cite{sm}].

To construct the  spin Hamiltonian  of NiI$_2$, we start  with distant neighbors going  up to fifth nearest neighbors, spin-orbit coupling (SOC) effects, and up to four-body interactions (see SM \cite{sm}).
Energies of random spin structures are calculated with HSE06 hybrid functional \cite{krukau2006influence}, but also with PBE+U for comparison \cite{perdew1996generalized}.
After repeated fitting and refining, the model finally reads
\begin{equation}
\begin{aligned}
\mathcal{H}= &\sum_{\langle i,j \rangle_1}\{J\bm{{\rm S}}_i{\cdot}\bm{{\rm S}}_j + KS_i^{\gamma}S_j^{\gamma} +
 B(\bm{{\rm S}}_i{\cdot}\bm{{\rm S}}_j)^2 \} \\
 + &\sum_{\langle i,j \rangle_{n}}J_n\bm{{\rm S}}_i{\cdot}\bm{{\rm S}}_j +  \sum_{\langle i,j \rangle_{n}^\perp}J_n^\perp\bm{{\rm S}}_i{\cdot}\bm{{\rm S}}_j + \sum_{i} A_{zz}{\rm S}^z_i{\rm S}^z_i
\end{aligned}
\end{equation}
with $n=1,2,3$ and where $\langle i,j\rangle_n$ denotes pairs of $n$th nearest neighbors (NN) within each layer, while the $\perp$ symbol refers to interlayer couplings; $\gamma$ chooses its value from $X,Y$ and $Z$ from the Kitaev basis (see Fig. 1d and SM \cite{sm}), which shows the bond-dependent feature.
Note that the SOC effects are reflected by the Kitaev term and SIA.
For the sum running over $\langle i,j\rangle_1$, $J$ quantifies the isotropic exchange coupling, $K$ the Kitaev interaction, and $B$ a biquadratic term. Note that one can also define $J_1=\frac{1}{3}(3J+K)$, which can be thought of as the real isotropic exchange.
$A_{zz}$ denotes the SIA.
As shown in Table I, the 1NN isotropic exchange favors FM since $J=-4.976$ meV, which is the largest coefficient in magnitude. $J_2$ also favors FM because of its negative sign, but is relatively very small. On the other hand, $J_3=2.250$ meV favors AFM and thus competes with the 1NN $J$.
Regarding the interlayer couplings, $J_1^{\perp}$ is FM in nature but very small in magnitude. In contrast, $J_2^\perp=0.685$ meV favors AFM and is the strongest interlayer coupling.
Moreover, sizable AFM Kitaev $K=0.858$ meV  and $B=-0.719$ meV biquadratic interactions  are predicted, which are in line with previous studies \cite{stavropoulos2019microscopic,amoroso2020spontaneous,ni2021giant,riedl2022microscopic,xu2018interplay,xu2020possible}. Such spin Hamiltonian of Eq. (1) yields a very small mean averaged error (MAE) of 0.063 meV/Ni, as indicated in the SM \cite{sm}.

\begin{table}[tbp]\centering
  \caption{Magnetic parameters of Eq. (1) fitted from different DFT functionals, as well as their ratio with $J$ shown in parentheses, in unit of meV. The $\perp$ symbol denotes interlayer couplings.}
  \renewcommand\arraystretch{1.4}
  \begin{tabular}{crrrr}
  \hline\hline
  \hspace{2mm} NiI$_2$ \hspace{2mm}  & \multicolumn{2}{c}{HSE} & \multicolumn{2}{c}{PBE}    \\
  \hline
  $A_{zz}$     &\hspace{4mm}  0.140   &\hspace{3mm} (-0.03)~~~~~ &\hspace{4mm}  0.212    &\hspace{3mm} (-0.05)~~~~~    \\
  $J$          &\hspace{4mm} -4.976   &\hspace{3mm} ( 1.00)~~~~~ &\hspace{4mm} -4.338    &\hspace{3mm} ( 1.00)~~~~~    \\
  $K$          &\hspace{4mm}  0.858   &\hspace{3mm} (-0.17)~~~~~ &\hspace{4mm}  1.433    &\hspace{3mm} (-0.33)~~~~~    \\
  $B$          &\hspace{4mm} -0.719   &\hspace{3mm} ( 0.14)~~~~~ &\hspace{4mm} -0.685    &\hspace{3mm} ( 0.16)~~~~~    \\
  $J_2$        &\hspace{4mm} -0.155   &\hspace{3mm} ( 0.03)~~~~~ &\hspace{4mm} -0.121    &\hspace{3mm} ( 0.03)~~~~~    \\
  $J_3$        &\hspace{4mm}  2.250   &\hspace{3mm} (-0.45)~~~~~ &\hspace{4mm}  3.155    &\hspace{3mm} (-0.73)~~~~~    \\
  $J_1^\perp$  &\hspace{4mm} -0.048   &\hspace{3mm} ( 0.01)~~~~~ &\hspace{4mm} -0.060    &\hspace{3mm} ( 0.01)~~~~~    \\
  $J_2^\perp$  &\hspace{4mm}  0.685   &\hspace{3mm} (-0.14)~~~~~ &\hspace{4mm}  1.103    &\hspace{3mm} (-0.25)~~~~~    \\
  $J_3^\perp$  &\hspace{4mm}  0.105   &\hspace{3mm} (-0.02)~~~~~ &\hspace{4mm}  0.195    &\hspace{3mm} (-0.04)~~~~~    \\
  \hline\hline
  \end{tabular}
\end{table}

The ground state of NiI$_2$ is determined employing the Hamiltonian of Eq. (1) within MC and conjugate gradient (CG) methods. The predicted ground state indeed yields a canted PS state with an in-plane $\langle1\bar10\rangle$ propagation and antiparallel interlayer alignments, which agree well with measurements. The period is determined to be $\lambda=7.3{\bm a}$ if neglecting interlayer couplings, which compares well with the experimental value of $\lambda=7.23{\bm a}$ (where ${\bm a}$ denotes the in-plane lattice constant)  \cite{kuindersma1981magnetic,kurumaji2013magnetoelectric}.
Strikingly the canting angle of the rotation plane is numerically found to be 46$^\circ$ for bulk, which is consistent with the corresponding measured  value of 55$^\circ$$\pm$10$^\circ$  \cite{kuindersma1981magnetic}. Our model therefore  reproduces well the correct PS state for bulk, where the spin texture in a single layer will be referred to as PS$^{\langle1\bar10\rangle}_{cant}$ state.
Note that the  parameters from PBE result in the $\langle110\rangle$ propagation, as a result of rather strong $J_3/J$.
It is also important to know  that isotropic Heisenberg terms, by themselves, do not support in-plane $\langle1\bar10\rangle$ propagation, as
$J_2$ and $J$ both favor FM while $J_3$ and $J$ compete against each other (since $J_3>0$ and thus favor AFM while $J_3/J=-0.45$).
Such isotropic Heisenberg terms lead to an incommensurate state along $\langle110\rangle$ (IC$^{\langle110\rangle}$), which is consistent with both analytical results \cite{okubo2012multiple} and previous models from DFT \cite{amoroso2020spontaneous,riedl2022microscopic,ni2021giant,ni2022plane}.
It therefore indicates that the $\langle1\bar10\rangle$ propagation is stabilized by mechanisms other than the isotropic Heisenberg terms.

\begin{figure}[btp]
  \centering
  \includegraphics[width=8.5cm]{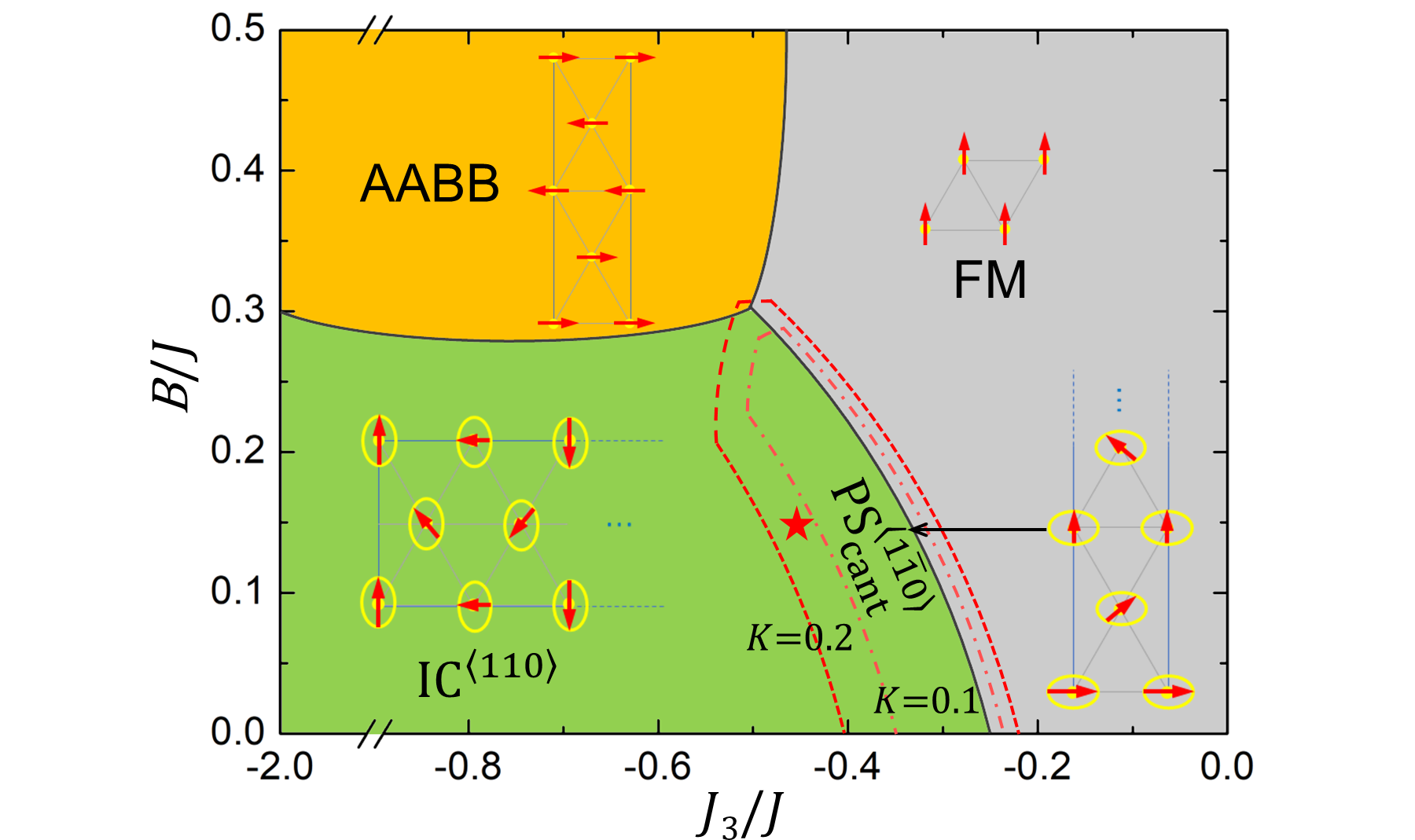}%
  \caption{Phase diagram for the studied triangular lattice. $J=-1$ meV is fixed in these calculations, $J_3$ and $B$ can vary in magnitude but not in sign. The red dashed (respectively, dot-dashed) line indicates that the ground state becomes IC$^{\langle1\bar10\rangle}$ (more precisely, PS$^{\langle1\bar10\rangle}_{cant}$) when $K/J=-0.2$ (respectively, $K/J=-0.1$). The red star denotes the model-predicted position in this phase diagram for NiI$_2$. Note that this phase diagram is determined by initial MC simulations and further CG optimizations, which guarantee its accuracy (see SM for details \cite{sm}).}
\end{figure}

To unravel the puzzling mechanisms that stabilize such $\langle1\bar10\rangle$ propagation, we built a phase diagram.
More precisely, we chose $J=-1$ meV and sweep over $J_3$ $\ge$ 0 and $B$ $\le$ 0 (in this phase diagram,  ``only'' $J$, $J_3$ and $B$ are thus included for now). As shown in Fig. 2, for $B/J=0$, the chosen negative $J$ stabilizes the FM state when $J_3$ is weak; while the system adopts IC$^{\langle110\rangle}$ states when $J_3/J<-0.25$, which is consistent with the analytical results of Ref.\cite{okubo2012multiple}.
For $B/J>0$, the negative biquadratic term shifts the IC$^{\langle110\rangle}$-FM boundary toward larger magnitude of $J_3/J$, which can be understood by the fact that $B<0$ favors collinear arrangements and thus helps stabilize the FM state.
When $B/J\gtrsim0.3$ and $J_3/J\lesssim-0.5$, a so-called AABB AFM state becomes the ground state \cite{wu2022plane}.
Moreover, calculations varying the interlayer Heisenberg terms ($J_n^{\perp}$, $n=1,2,3$) were also performed, but their results are not shown in the phase diagram. It is found that $J_n^{\perp}$ can only modify the period of IC states or induce collinear states, but not alter the propagation direction, which is in line with a previous work too \cite{regnault1982inelastic}.
The phase diagram discussed so far thus indicates that $J$, $J_3$, $B$ and $J_n^{\perp}$, by themselves, can not lead to the $\langle1\bar10\rangle$ propagation in the investigated parameter space.

\begin{table}[t]\centering
  \caption{Total energy and relative energies of different PS states, as well as the decomposition of these energies into specific interaction, as calculated with the HSE parameters in Table I.  (unit: meV/Ni). }
  \renewcommand\arraystretch{1.4}
  \begin{tabular}{>{\hfil}p{22pt}>{\hfil}p{30pt}>{\hfil}p{30pt}>{\hfil}p{30pt}>{\hfil}p{35pt}>{\hfil}p{35pt}>{\hfil}p{35pt}}
  \hline\hline
  \multirow{2}{*}{Para.}   & \multirow{2}{*}{PS$^{\langle1\bar10\rangle}_{cant}$} & \multirow{2}{*}{PS$^{\langle1\bar10\rangle}$} & \multirow{2}{*}{PS$^{\langle110\rangle}$}       & PS$^{\langle1\bar10\rangle}$  & PS$^{\langle1\bar10\rangle}_{cant}$  & PS$^{\langle1\bar10\rangle}_{cant}$  \\
   & & & & -PS$^{\langle110\rangle}$ & -PS$^{\langle110\rangle}$ & -PS$^{\langle1\bar10\rangle}$ \\
  \hline
  $A_{zz}$    &     ~0.04      &    ~0.07   &       ~0.07 &         ~0.00   &       -0.03   &       -0.03     \\ 
    $J$       &    -11.42      &   -11.42   &      -11.39 &         -0.03   &       -0.03   &       ~0.00     \\ 
    $K$       &     ~0.56      &    ~0.61   &       ~0.61 &         ~0.00   &       -0.05   &       -0.05     \\ 
    $B$       &     ~0.84      &    ~0.84   &       ~0.85 &         -0.01   &       -0.01   &       -0.00     \\ 
    $J_2$     &     -0.18      &    -0.18   &       -0.17 &         -0.00   &       -0.00   &       ~0.00     \\ 
    $J_3$     &     ~1.52      &    ~1.52   &       ~1.45 &         ~0.07   &       ~0.07   &       ~0.00     \\ 
    Total     &     -8.64      &    -8.56   &       -8.59 &         ~0.03   &       -0.05   &       -0.08     \\ 
  \hline\hline
  \end{tabular}
\end{table}


The Kitaev interaction is therefore now further incorporated into the computations and resulting phase diagram (consequently,  $J$, $J_3$, $B$ and $K$ are now included in this new phase diagram). Surprisingly, with $K=0.1$ meV (resulting thus in $K/J=-0.1$),  an incommensurate state propagating along $\langle1\bar10\rangle$ (IC$^{\langle1\bar10\rangle}$) emerges at the border of the previous IC$^{\langle110\rangle}$-FM transition, as additionally shown in Fig. 2.
Such IC$^{\langle1\bar10\rangle}$ state takes a slim area of the previous FM zone and a relatively large area of the previous IC$^{\langle110\rangle}$ state. When increasing the Kitaev interaction even more to $K=0.2$ meV, the area of IC$^{\langle1\bar10\rangle}$ state further expands.
As a result, the phase points defined by, {\it  e.g.}, $J_3/J=-0.4$ and $B/J=0$, as well as $J_3/J=-0.5$ and $B/J=0.2$, transform from the IC$^{\langle110\rangle}$  to IC$^{\langle1\bar10\rangle}$ state.
It is thus clear that, for NiI$_2$, the ratios $J_3/J=-0.45$ and $B/J=0.14$ favor the IC$^{\langle110\rangle}$ state, but $K/J=-0.17$ renders the ground state to become the IC$^{\langle1\bar10\rangle}$ state.
Such results therefore demonstrate that the Kitaev interaction (with $K>0$), along with the frustration between $J$ and $J_3$,   tends to stabilize the ${\langle1\bar10\rangle}$ propagation.

\begin{figure}[t]
  \centering
  \includegraphics[width=8cm]{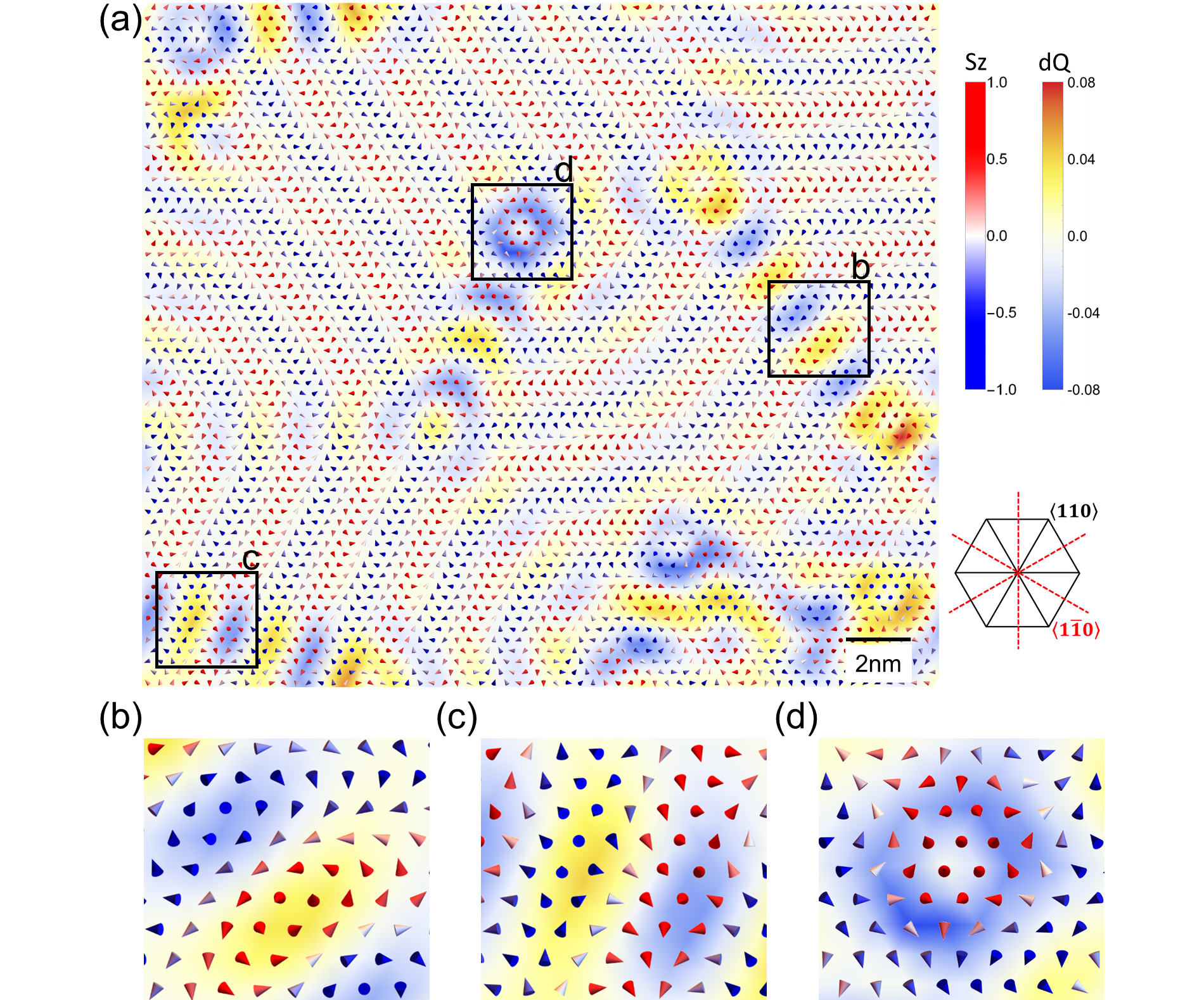}%
  \caption{Panel (a) displays spin patterns of NiI$_2$ from MC simulation and a following CG optimization \cite{note2}. Panels (b), (c) and (d), respectively, show a zoom-in view of the topological defects occurring in Panel (a). Spins are represented by cones, with the red and blue colors showing positive and negative values of the $S_z$ component. For all these panels, the colors used for the background quantify the topological charges, $dQ$. }
\end{figure}

Moreover, it is found that the aforementioned IC$^{\langle1\bar10\rangle}$ state resulted from the $J$-$K$-$J_3$(-$B$) model (i.e. a model with only such terms) also exhibits canted rotation plane.
This canting angle between the $Y$ axis and out-of-plane direction yields 54.7$^\circ$, implying that the canting plane locates exactly in the $XZ$ plane (see Fig. 1d for the Kitaev basis).
If we focus on a single layer, such canted IC$^{\langle1\bar10\rangle}$ state is actually the PS$^{\langle1\bar10\rangle}_{cant}$ state of NiI$_2$.
Such fact strongly suggests that the canting of the PS$^{\langle1\bar10\rangle}_{cant}$ state for NiI$_2$ is strongly related to the Kitaev interaction. To verify such point, we turned on only the intralayer terms of Eq. (1) and compare the energies of three phases: (i) PS$^{\langle1\bar10\rangle}_{cant}$ state with a period of $\lambda=7.25{\bm a}$; (ii) artificially made PS$^{\langle1\bar10\rangle}$ state (that has no canting) with $\lambda=7.25{\bm a}$; and (iii) artificially made PS$^{\langle110\rangle}$ state (that has also no canting) with $\lambda=6.25{\bm a}$.
Note that the chosen periods lead to the lowest energy of the corresponding propagation.
The total and decomposed energies of such three phases are listed in Table II.
It is found that, (a) PS$^{\langle1\bar10\rangle}$ is 0.03 meV/Ni higher in energy than PS$^{\langle110\rangle}$ and the Kitaev term contributes the same energy to both states, indicating that the Kitaev interaction favor PS$^{\langle1\bar10\rangle}$ and PS$^{\langle110\rangle}$ equally;
while (b) PS$^{\langle1\bar10\rangle}_{cant}$ is 0.08 meV/Ni lower in energy than PS$^{\langle1\bar10\rangle}$, and the Kitaev interaction, as well as the SIA, contributes dominantly to this energy gain.
Such comparisons thus demonstrate that the Kitaev interaction favors canting altogether with the ${\langle1\bar10\rangle}$ propagation.
Moreover, it indicates that Kitaev interaction favors spins rotating in $ZX$, $XY$ or $YZ$ planes with a canting angle of 54.7$^\circ$, while the in-plane SIA further pushes the rotation plane toward the basal plane with a canting angle of 46$^\circ$.

We then develop a model to better understand why Kitaev interaction favors ${\langle1\bar10\rangle}$ propagation, as well as,  a canting in rotation plane (see details in SM \cite{sm}).
Here, we construct PS$^{\langle1\bar10\rangle}$ and PS$^{\langle110\rangle}$ states and adopt only the Kitaev interaction.
The resulted energies are expressed as $E^{1\bar10}/K=c_1({\rm cos}2\theta_1-2\sqrt{2}{\rm sin}2\theta_1)+c_2$ and $E^{110}/K=c_3 {\rm cos}2\theta_1+c_4$, where $E^{1\bar10}$ and $E^{110}$ are the total energies of the corresponding PS states, $\theta_1$ is the angle from the [001] direction to the normal of the rotation plane, and $c_n (n=1-4)$ are positive constants.
It is found that (i) $E^{1\bar10}$ has its minimum at $\theta_1=54.7^{\circ}$, which is the angle between the [001] direction and the $Y$ ($Z$ or $X$, respectively) axis, demonstrating that the Kitaev interaction prefers the rotation plane of the PS$^{\langle1\bar10\rangle}$ pattern within the $XZ$ ($XY$ or $YZ$, respectively) plane; (ii) $E^{110}$ has its minimum at $\theta_1=\pm90^{\circ}$, indicating an exact PS state with rotation plane being perpendicular to the propagation direction; and (iii) $E^{1\bar10}_{min}<E^{110}_{min}$, confirming that ${\langle1\bar10\rangle}$ propagation, together with a canting, is energetically more  favorable (see Fig. S3 of SM \cite{sm}).

The critical role of Kitaev interaction in reproducing the canting in spin rotation plane demonstrates the significance of SOC effects on the spin model of NiI$_2$. Moreover, our DFT results (see Fig. S7 of SM) show that the strength of electric polarization depends largely on the orientation of the spin rotation plane. It thus indicates that the Kitaev interaction is closely related to the ferroelectricity. Such findings are thus in line with previous work, which demonstrates that the ferroelectric order is controlled by the SOC of iodine \cite{fumega2022microscopic}.

Furthermore, Monte-Carlo simulations, as well as a conjugate gradient (CG) algorithm, are performed on large supercells using the Hamiltonian  of Eq. (1).
Since bulk only differs from the monolayer only by a longer period of propagation and interlayer AFM alignments, we focus on the monolayer hereafter for simplicity. As shown in Fig. 3(a), these simulations found that canted PS states form stripy domains and cover most of the area at low temperatures, which is consistent with the fact that the PS$^{\langle1\bar10\rangle}_{cant}$ states are the ground states of NiI$_2$ bulk.
There are three domains that propagate along $\langle1\bar10\rangle$ or the equivalent $\langle120\rangle$ and $\langle\bar2\bar10\rangle$ directions, which is also in line with the observed three domains of NiI$_2$ monolayer \cite{song2022evidence}. Note that the spin pattern shown in Fig. 3a is only 0.038 meV/Ni higher in energy than the ground state of PS$^{\langle1\bar10\rangle}_{cant}$ monodomain.
Interestingly, topological defects are predicted to occur at phase boundaries (see Fig. 3), which is in line with the prediction of skyrmion lattice in monolayer NiI$_2$ \cite{amoroso2020spontaneous}.

To conclude, we adopted the symmetry-adapted cluster expansion method and built a realistic spin model for multiferroic NiI$_2$. Such model can reproduce well the experimental $\langle1\bar10\rangle$ propagating proper screw state, as well as the canting in its spin rotation plane. The Kitaev interaction is found to play a key role in NiI$_2$, and is proved to impose anisotropy on coplanar spin texture. Our work thus leads to a better understanding on the magnetism of NiI$_2$, as well as its type-II multiferroicity.

\begin{acknowledgments}
This work is supported by NSFC (grants No. 811825403, 11991061, 12188101, 12174060, and 12274082) and the Guangdong Major Project of the Basic and Applied Basic Research (Future functional materials under extreme conditions--2021B0301030005). C.X. also acknowledge support from the the support
from Shanghai Science and Technology Committee (grant No. 23ZR1406600) and the open project of Guangdong provincial key laboratory of magnetoelectric physics and devices (No. 2020B1212060030).
L.B. acknowledges support from the Vannevar Bush Faculty Fellowship (VBFF)
from the Department of Defense and the ARO Grant No. W911NF- 352-21-2-0162 (ETHOS).
The Arkansas High Performance Computing Center (AHPCC) is also acknowledged.
\end{acknowledgments}



\begin{thebibliography}{34}
\bibitem [{not()}]{note2}%
  \BibitemOpen
  \href@noop {} {\ }\bibinfo {note} {Note that the spin patterns in Fig. 3(a)
  do not correspond to a MC temperature, as the CG algorithm leads the pattern
  to a stable state, which is a global or local energy minimum.}\BibitemShut
  {Stop}%
\end{thebibliography}
%

\end{document}